\documentstyle[12pt,aasms4]{article}

\received{}
\accepted{}
\journalid{}{}
\articleid{}{}

\lefthead{Brunner et al.}
\righthead{Photometric-Redshifts}

\newcommand{\UBRI}{{$U, B, R, \&\ I \ $}}
\newcommand{\ubri}{{\em U B R I\ }}
\newcommand{\ubr}{{\em U B R\ }}
\newcommand{\bri}{{\em B R I\ }}

\newcommand{\HST}{{\it HST\ }}

\newcommand{\eg}{{\em eg.\ }}
\newcommand{\cf}{{\em cf.\ }}
\newcommand{\etal}{{\em et al.\ }}

\begin{document}

\title{Towards More Precise Photometric Redshifts: Calibration Via CCD Photometry}

\author{Robert J. Brunner\altaffilmark{1,2}, 
Andrew J. Connolly\altaffilmark{1}, 
and Alexander S. Szalay\altaffilmark{1}}
\affil{rbrunner@pha.jhu.edu, ajc@pha.jhu.edu, szalay@pha.jhu.edu\\
Department of Physics \& Astronomy, The Johns Hopkins University, 
Baltimore, MD 21218}

\and

\author{Matthew A. Bershady\altaffilmark{1}}
\affil{
mab@astro.psu.edu\\
Department of Astronomy \& Astrophysics,
Penn State University,
University Park, PA 16802}

\altaffiltext{1}{Visiting Astronomer, Kitt Peak National Observatory, National
Optical Astronomy Observatories, which is operated by the
Association of Universities for Research in Astronomy, Inc. (AURA)
under cooperative agreement with the National Science Foundation.}
\altaffiltext{2}{NASA GSRP Fellow}

\begin{abstract}

We present the initial results from a deep, multi-band photometric
survey of selected high Galactic latitude redshift fields.  Previous
work using the photographic data of Koo and Kron demonstrated that the
distribution of galaxies in the multi-dimensional flux space \UBRI is
nearly planar. The position of a galaxy within this plane is
determined by its redshift, luminosity and spectral type
(\cite{connolly95}). Using recently acquired deep CCD photometry in
existing, published redshift fields, we have redetermined the
distribution of galaxies in this four-dimensional magnitude
space. Furthermore, from our CCD photometry and the published
redshifts, we have quantified the photometric-redshift relation within
the standard AB magnitude system. This empirical relation has a
measured dispersion of $\sigma_{z} \approx 0.02$ for $z < 0.4$. With
this work we are reaching the asymptotic intrinsic dispersions
($\sigma_{z} \approx 0.016$ for $z < 0.4$) that were predicted from
simulated distributions of galaxy colors. This result will prove
useful in providing estimated redshifts for large photometric surveys,
as well as improve the sampling of specific redshift regions for
spectroscopic surveys through the use of an estimated redshift
selection criteria.

\end{abstract}

\keywords{cosmology: observations - galaxies: evolution - galaxies: photometry}

\section{Introduction}

The utility of deriving galaxy redshifts from photometric data has
long been known (\cite{baum62,koo85,loh86}). Recently,
\cite{connolly95} developed an empirical approach as opposed to the
previous model fitting methods. Utilizing photographic data, they were
able to estimate a redshift out to $z \sim 0.5$ with a measured
dispersion of $\delta z < 0.05$. The uncertainties in that result
were dominated by the photometric errors. Simulations indicated
that with improved photometry the dispersion within the relationship
could be significantly reduced. As a result, we have embarked on an
observational program designed to obtain deep CCD multi-band
photometry in existing redshift fields.

In this paper we present the first results of this survey by extending
the previous approach using CCD photometry. Section two outlines the
basic reduction steps taken in the preparation of the sample for this
work. Section three discusses the actual fitting techniques and
investigates the intrinsic dispersion. We conclude this letter with a
discussion of the ramifications of this work and possible future
directions.

\section{Data}

The photometric data used in this analysis were taken using the PFCCD
camera with the standard \UBRI filters on the Mayall 4m at Kitt Peak
National Observatory on the nights of March 31 $-$ April 3, 1995,
March 18$-$20, 1996, and May 14$-$16, 1996. This camera uses a
$2048^2$ CCD ({\em T2KB}) with a $0.47\arcsec$/pixel
scale and a read noise of $4 e^{-}$/pixel. The gain used for these
observations was $5.4\  e^{-}/ADU$, a value which resulted from a
tradeoff between maximizing the available dynamic range and minimizing
the effects of the charge depletion problem with the CCD
electronics. These observations were chosen to coincide with the
published 14 hour redshift field of the Canada-France Redshift Survey
(CFRS). A complete discussion of the observational program including
an analysis of the custom reduction software are beyond the scope of
this letter and will be published elsewhere (\cite{myThesis}).

All three runs were reduced separately using the standard IRAF
routines. The images were initially debiased and flat fielded using
dome flats. Illumination corrections were created by stacking the
image frames in each filter, with high and low rejection to remove
objects, and then boxcar smoothing the stacked image. The individual
fields were registered to a common position in each filter, and then
stacked using a weighted average. The weights were determined by
measuring the signal to noise for several randomly chosen stars on
each frame. The stacked images for each filter were then registered to
a common image to simplify the photometric measurement in matched
apertures. The final images for each of the three runs were then
registered and stacked again using the weighted average.

Object detection and photometry were performed using a custom
pipeline. The object detection was done separately in each filter
using the SExtractor package (\cite{bertin96}). The separate
detections in each filter were then matched using a growing annulus
technique in the order of $B$, $U$, $R$, then $I$ and a master
detection list was produced. Using this list, photometry was
determined in both SExtractor's modified Kron (\cite{kron80}) aperture
and a $10 \arcsec$ diameter aperture matched in each band. The actual
photometry algorithm used involved a modification to SExtractor in
both the background calculation and pixel assignment within the
aperture of interest. The detections were photometrically calibrated
using published standards (\cite{landolt92}) which were measured at
similar airmasses to the object frames. The photometric zeropoint was
then adjusted to the AB system (\cite{okeGunn83}) using published
transformations (\cite{fukugita95}): $U_{AB} = U + 0.69$, $B_{AB} = B
- 0.14$, $R_{AB} = R + 0.17$, and $I_{AB} = I + 0.44$.

Astrometric transformations were determined from the \HST Guide Star
Catalog II (\cite{gsc2}), after which the redshifts in the Canada
France Redshift Survey (CFRS) 14 hour field were matched to our
photometric sample. The measured dispersion between the CFRS $I_{AB}$
isophotal magnitudes and our $I_{AB}$ automatic magnitudes was $\sigma
\approx 0.13$ to $I \sim 23$ with no evident systematic deviations
from a linear relationship.

\section{Analysis}

In an effort to minimize the dispersion in our relationship, we
imposed two conditions on the data used in this analysis. First, we
restricted the photometric sample such that all object magnitudes were
below the appropriate magnitude limit at which a typical galaxy had a
measured $1 \sigma_{rms}$ magnitude error of approximately $0.1$
magnitudes. Second, we required that only the most reliable
spectroscopic identifications be incorporated into the fitting
procedure. This involved pruning the CFRS catalog such that only
non-stellar objects with redshifts having a confidence greater than
95\% were retained. This was accomplished by restricting the redshifts
used to the following quality classes: 3,4,8,93,94,98 (\cf
\cite{CFRS2}).

The final sample contained 89 redshifts with the following
distribution: 40 redshifts in the range (0.0, 0.4] and 49 redshifts in
the range (0.4, 0.8]. For these two subsets, we fit a second order
polynomial in \ubri, \ubr, and \bri to the measured photometry and
published redshifts. In each region, the degrees of freedom remained a
substantial fraction of the original data (a second order fit in four
variables requires 15 parameters). This technique is a simple approach
designed to quantify the accuracy of our method for estimating
redshifts and is not the optimal parameterization of the topology of
the galaxy distribution, which is the subject of ongoing work.

\placetable{TableOne}

The redshift intervals were not chosen randomly. This technique has
been previously shown to be more sensitive to broad spectroscopic
continuum features (primarily the break in the continuum spectra at
around 4000 \AA, which moves between the $B$ and $R$ bands at $z
\approx 0.4$, and the $R$ and $I$ bands at $z \approx 0.8$) rather
than specific absorption/emission features (\cite{connolly95}). This
is clearly demonstrated in Table \ref{TableOne} where the standard
deviation in the redshift range (0.0,0.4] is only slightly higher when
the $I$ band is not included in the fit. On the other hand, when the
$U$ band is excluded from the fit, the standard deviation more than
doubles. This reflects the fact that the $I$ band is sampling the same
flat region of the spectrum as the $R$ band within this redshift
range, and is thereby providing predominantly redundant information to
the fit. In the second redshift range, the continuum break is moving
from the $B$ band into the $R$ band, which is reflected in the lower
significance of the $U$ band information. We also show the expected
intrinsic dispersion in this relationship from simulations using all
four bands (\cf \cite{connolly95}), which clearly shows that our
measured dispersion is completely dominated by the intrinsic scatter
within the relationship.

\placefigure{FigureOne}

The relative importance of the different bands in the individual
redshift intervals reflects the curvature inherent within the
distribution of galaxies in the multi-dimensional magnitude space. In
a given redshift range, the curvature is accurately parameterized by a
second order polynomial. Between redshift intervals, however, the
distribution displays a higher order curvature term (\cf the previous
discussion concerning the continuum break), which requires
the use of a piecewise second-order parameterization. As a result, the
application of these results requires an iterative approach. First, a
third order global photometric redshift relation is used to determine
an approximate redshift. From this initial redshift estimate, the
appropriate second-order relationship can then be used. If the initial
estimate is on the border between two subsets ($z \subset [0.35,0.45]$),
both relationships should be applied and the mean of the two results
used.

With the introduction of the four-vector $C = (U,B,R,I)$, the
second-order photometric-redshift relationships can be summarized
in the following manner:
\begin{displaymath}
z = Z_{\alpha} + C \cdot V_{\alpha} + C \cdot M_{\alpha} \cdot C^{T}
\end{displaymath}
where the scalar $Z_{\alpha}$, vector $V_{\alpha}$, and matrix
$M_{\alpha}$ components are listed in Table \ref{TableTwo} for the two
different redshift regimes. The parameters for the third order fit are
listed in Table \ref{TableThree}.

\section{Discussion}

We have shown that using a simple iterative process, redshifts can be
reliably estimated for objects from broadband photometry out to $z
\sim 0.8$. A comparison of our measured dispersion with the published
intrinsic dispersion from simulations (\cite{connolly95}) indicates
that we have approached the inherent scatter within the
photometric-redshift relationship. These simulations provide an
absolute lower limit to the intrinsic scatter, as they only assumed an
evolved (15 Gigayear) SED. As the additional effects of metallicity,
dust, and stellar histories can only increase the scatter within the
relationship, we do not include them in our estimation of the minimal
intrinsic scatter within the photometric-redshift relation. 

Thus it is quite remarkable that we measure such a small scatter as
compared to the simple, single age, solar metallicity, and dust free
galaxies produced in the simulations. Actual galaxies are clearly more
complex, spanning a wide range of star formation histories, ages,
metallicities, and dust content, all of which would be expected to
significantly increase the measured dispersion. We see that this is
not the case, which leads us to two related conclusions. First, this
technique is extremely dependent on the $4000$ \AA\ break which is
present in nearly all galaxies. Second, metallicity, dust, and age
variations have similar effects in this multidimensional space, albeit
almost orthogonal to the redshift vector (\cite{koo86}). We plan on
improving our understanding of the multidimensional nature of the
observed galaxy population through the use of SED modeling. This will
allow us to quantify the importance of the metallicity, dust, and
different stellar histories and explore any possible degeneracies.

The photometric-redshift estimation technique can be considered to be
the equivalent of a low resolution (4 element) spectrograph. By using
more filters that are increasingly narrow, we increase the spectral
resolution of this technique. Taken to the extreme, however, this
approach will emulate a spectrograph, losing the observing efficiency
that is the primary advantage of this technique. From a comparison
between the dispersion from the three band and the four band quadratic
fits, it is clear that a marginal gain is achieved by adding a fourth
band within a given redshift regime. As a result, we believe that the
benefits achieved by adding more bands to this approach is more than
offset in the loss of observational efficiency. The simulations used
the standard \UBRI filters in order to be reliably compared to our
observations.

We are working to extend this analysis in two principal areas. First,
we are now focusing on improving our understanding of the distribution
of galaxies in this multi-dimensional flux space. This requires the
use of a stratified sampling strategy to obtain redshifts throughout
the photometric sample. These additional redshifts are primarily being
obtained using the Keck telescope within the DEEP collaboration. In
addition, we are incorporating additional physical parameters (\eg
surface brightness and shape parameters) via \HST WFPC2 imaging to
quantify the different morphological tracts within the cumulative
galaxy distribution.

Second, we are extending this work to higher redshifts.  For redshifts
below $z \sim 1.2$, we are working to increase the size and
stratification of our redshift sample. This involves increasing our
photometry-redshift catalog through the addition of published
redshifts and our participation in several spectroscopic surveys. In
the redshift region $1.2 \leq z \leq 2.8$, we are adding near-infrared
photometry to our catalog in order to sample the continuum features
our technique requires. Until a large quantity of reliable spectra can
be obtained within this region (the arrival of the blue camera on LRIS
will help alleviate this quandary), we will use our understanding of
the $z < 1.2$ regime and the published high $z$ work of others
(\cite{steidel96}) as boundary conditions. We can then use SED models
to extrapolate into this region, while maintaining the boundary
condition requirements at both redshift ends. As spectra in this area
become available, we will add them into the fitting procedure.
Eventually this work will allow for the estimation of not only the
redshift, but also the spectral type of an object solely from
broadband photometry.

\acknowledgments

First we wish to acknowledge Gyula Szokoly for assistance in obtaining
the data. We also would like to thank the referee for useful comments,
and Barry Lasker, Gretchen Greene, and Brian McLean for allowing us
access to an early version of the GSC II. We also wish to acknowledge
useful discussions with Mark Dickinson, Mark Subbarao, and David
Koo. RJB would like to acknowledge support from the National
Aeronautics and Space Administration Graduate Student Researchers
Program. AJC acknowledges partial support from NASA grant
AR-06394.01-95A. ASZ has been supported by the NASA LTSA program.

\clearpage

\begin{deluxetable}{ccccc}
\tablecaption{The standard deviation between measured and estimated redshifts.
\label{TableOne}}
\tablenum{1}
\tablewidth{0pt}
\tablehead{
\colhead{Redshift Range} 
& \colhead{$\sigma_{Z}(\ubri)$} 
& \colhead{$\sigma_{Z}(\ubr)$} 
& \colhead{$\sigma_{Z}(\bri)$}
& \colhead{$\sigma_{Z}(Simulation)$}
}
\startdata
(0.0,0.4]	&0.0234 	&0.0301		&0.0498	&0.016\nl
(0.4,0.8]	&0.0389		&0.0834		&0.0431 &0.043\nl
\enddata
\end{deluxetable}

\clearpage

\begin{deluxetable}{ccccccccc}
\tablecaption{The $2^{nd}$ order polynomial terms.\label{TableTwo}}
\tablenum{2}
\tablewidth{0pt}
\tablehead{
\colhead{Redshift Range} 
& \colhead{Scalar}
& \colhead{Vector}
& \colhead{Matrix}
& \colhead{}
& \colhead{}
& \colhead{}
& \colhead{}
\nl
\colhead{} 
& \colhead{}
& \colhead{}
& \colhead{U}
& \colhead{B}
& \colhead{R}
& \colhead{I}
& \colhead{}
}
\startdata
(0.0,0.4] &0.987	&-1.239  	&0.1168    &-0.25660 &-0.17239 &0.23456   &U\nl
	  &	 	&1.513  	&0.0	   &0.16295  &0.12049  &-0.23008  &B\nl
	  &	 	&2.099		&0.0	   &0.0	     &-0.22036 &0.39080   &R\nl
	  &	 	&-2.476		&0.0	   &0.0	     &0.0      &-0.14324  &I\nl
\nl		 				     				    
(0.4,0.8] &7.31  	&0.7245		&0.32111   &-1.1314  &0.83587  &-0.36212  &U\nl
	  &	 	&-2.493		&0.0       &0.97325  &-1.3183  &0.58085   &B\nl
	  &	 	&5.378 		&0.0	   &0.0      &0.52568  &-0.78022  &R\nl
	  &	 	&-4.225		&0.0       &0.0      &0.0      &0.36975   &I\nl
\enddata
\end{deluxetable}

\clearpage

\begin{deluxetable}{ccccccccc}
\tablecaption{The $3^{rd}$ order polynomial terms.\label{TableThree}}
\tablenum{3}
\tablewidth{0pt}
\tablehead{
\colhead{Scalar}
& \colhead{Vector}
& \colhead{Matrix}
& \colhead{}
& \colhead{}
& \colhead{}
& \colhead{}
\nl
\colhead{} 
& \colhead{}
& \colhead{U}
& \colhead{B}
& \colhead{R}
& \colhead{I}
& \colhead{}
}
\startdata
-31.5	&-3.209		&11.933    &-26.217  &-33.532  &35.985    &U\nl
 	&-18.34		&0.0	   &17.690   &38.390   &-46.131   &B\nl
 	&117.7		&0.0	   &0.0	     &-32.404  &50.319    &R\nl
 	&-90.97		&0.0	   &0.0	     &0.0      &-16.296   &I\nl
\tableline
\nl
&Third Order Terms 	&U	   &B	     &R	       &I	  &\nl
\nl
\tableline
26.391         &UBR     &-1.5113   &7.9136   &-8.0678  &4.1007    &U$^2$\nl
-12.859        &UBI     &-13.988   &8.4890   &-23.815  &11.322    &B$^2$\nl
-2.1809        &URI     &-3.3766   &13.023   &-5.5753  &8.2411    &R$^2$\nl
-6.0829        &BRI     &2.6542    &-0.91194 &-5.1008  &1.3397    &I$^2$\nl

\enddata
\end{deluxetable}

\clearpage

\figcaption[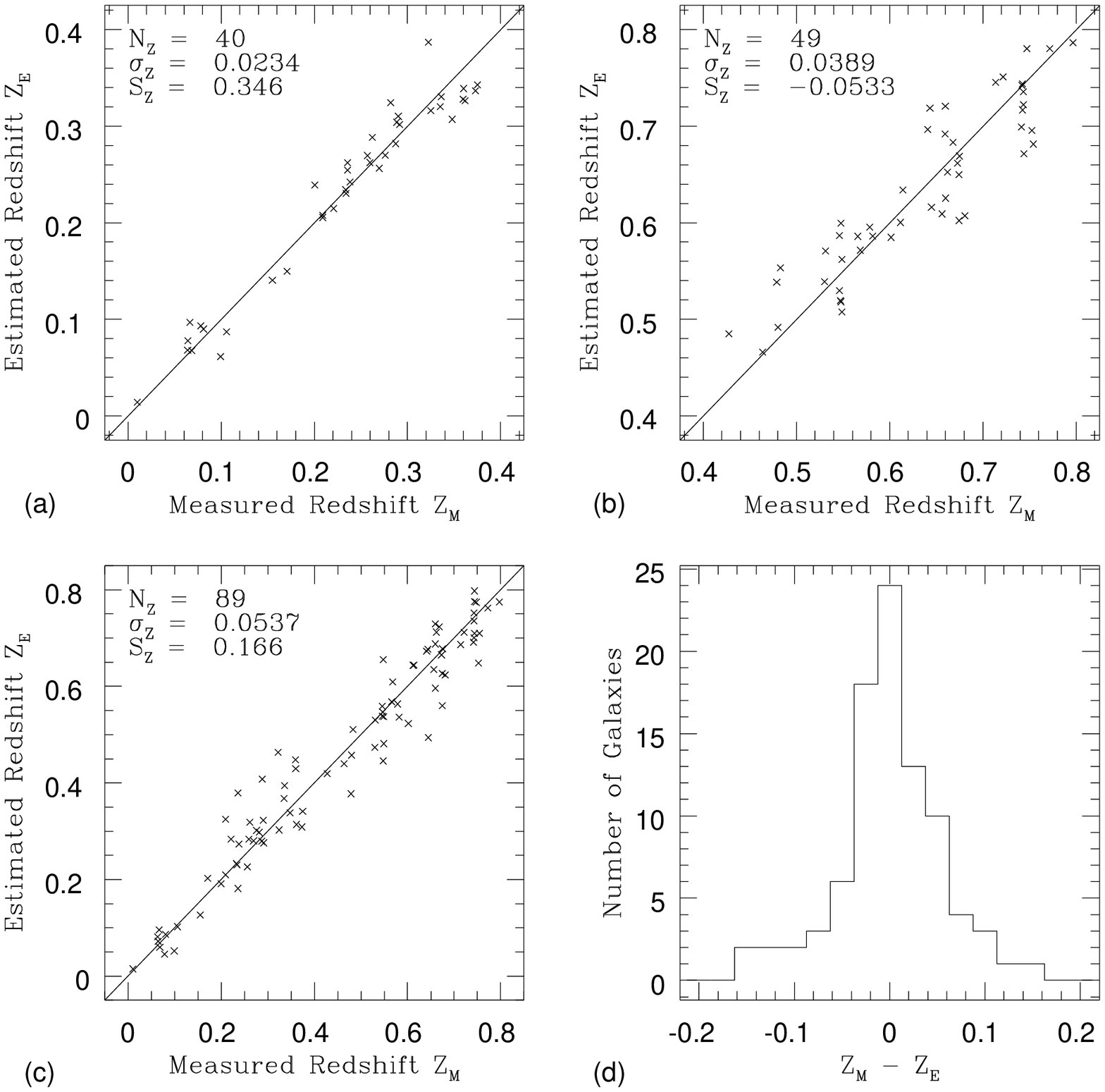]{The correlation between estimated and
measured redshifts are shown in (a) for the second order fit in the
redshift interval (0.0,0.4], (b) for the second order fit in the
redshift interval (0.4,0.8], and (c) for the third order fit in the
redshift interval (0.0,0.8]. Indicated in each figure are the number
of redshifts ($N_{Z}$), dispersion ($\sigma_{Z}$), skewness ($S_{Z}$),
and a straight line of unit slope, which is not a fit to the data. A
histogram of the residuals for the third order fit are displayed in
(d).\label{FigureOne}}

\clearpage

\plotone{figureOne.eps}

\end{document}